\documentclass[a4paper,fleqn,usenatbib]{mnras} 

\usepackage[T1]{fontenc}
\usepackage{ae,aecompl}

\usepackage{graphicx}	
\usepackage{amsmath}	
\usepackage{amssymb}	
\usepackage{multirow}

\title[Stellar wind-magnetosphere interaction at exoplanets]{Stellar wind-magnetosphere interaction at exoplanets: computations of auroral radio powers}

\author[J.~D.~Nichols and S. E. Milan ]{J.~D.~Nichols$^{1}$\thanks{E-mail:jdn@ion.le.ac.uk} and S. E. Milan$^{1}$\\
$^{1}$Department of Physics and Astronomy, University of Leicester, Leicester, LE1~7RH, UK}

\date{Accepted XXX. Received YYY; in original form ZZZ}

\pubyear{2016}

\begin{document}
\label{firstpage}
\pagerange{\pageref{firstpage}--\pageref{lastpage}}
\maketitle

\begin{abstract}
We present calculations of the auroral radio powers expected from exoplanets with magnetospheres driven by an Earth-like magnetospheric interaction with the solar wind.  Specifically, we compute the twin cell-vortical ionospheric flows, currents, and resulting radio powers resulting from a Dungey cycle process driven by dayside and nightside magnetic reconnection, as a function of planetary orbital distance and magnetic field strength.  We include saturation of the magnetospheric convection, as observed at the terrestrial magnetosphere, and we present power law approximations for the convection potentials, radio powers and spectral flux densities.  We specifically consider a solar-age system and a young (1~Gyr) system.  We show that the radio power increases with magnetic field strength for magnetospheres with saturated convection potential, and broadly decreases with increasing orbital distance.  We show that the magnetospheric convection at hot Jupiters will be saturated, and thus unable to dissipate the full available incident Poynting flux, such that the magnetic Radiometric Bode's Law (RBL) presents a substantial overestimation of the radio powers for hot Jupiters.  Our radio powers for hot Jupiters are $\sim$5-1300~TW for hot Jupiters with field strengths of 0.1-10~$B_J$ orbiting a Sun-like star, while we find that competing effects yield essentially identical powers for hot Jupiters orbiting a young Sun-like star.  However, in particular for planets with weaker magnetic fields our powers are higher at larger orbital distances than given by the RBL, and there are many configurations of planet that are expected to be detectable using SKA.   

\end{abstract}




\section{Introduction}

The radio waveband offers an extremely favourable contrast ratio for the direct detection of exoplanets, with e.g.\ Jupiter's non-thermal bursts as bright as the typical solar low frequency emissions \citep{zarka98a,zarka07b}. Interest in the radio emissions of exoplanets has further grown recently owing to commencement of observations of the Low Frequency Array (LOFAR), which has the potential to detect spectral flux densities of order $\simeq$1~mJy in 1 h integration at $\sim$10~MHz \citep{farrell04a}, and the imminent deployment of the Square Kilometer Array (SKA), which is expected to have a sensitivity of $\sim$10~$\mu$Jy in Phase 1 and $\sim$1~$\mu$Jy in Phase 2 \citep{Zarka:2015ui}.  Potentially-detectable exoplanetary radio emissions are envisaged to be excited by the electron cyclotron maser instability (CMI), the process responsible for generating coherent, powerful auroral radio emissions at the Earth and other planets in the solar system \citep{wu79a,2006A&ARv..13..229T}.  Attention has primarily focused on so-called `hot Jupiters' orbiting close to their parent star \citep[e.g.][]{farrell99a, farrell04a, zarka01a,zarka07a,lazio04a, griessmeier04a, griessmeier05a, griessmeier07a, stevens05a, jardine08a,Smith:2009fj, fares10a, reiners10a, vidotto11c,2011A&A...531A..29H,Saur:2013dc, 2015MNRAS.449.4117V, 2015MNRAS.450.4323S}, although \cite{nichols11a, nichols:2012aa} showed that further-orbiting, fast-rotating, massive planets orbiting XUV-bright stars are also capable of generating detectable radio emissions, and related emissions have possibly already been detected from fast-rotating ultra-cool dwarfs \citep{hallinan:2008aa, Berger:2010gf, mclean:2012aa, route:2012aa,nichols:2012b}.   \\ 

\begin{figure}
 \noindent\includegraphics[width=20pc]{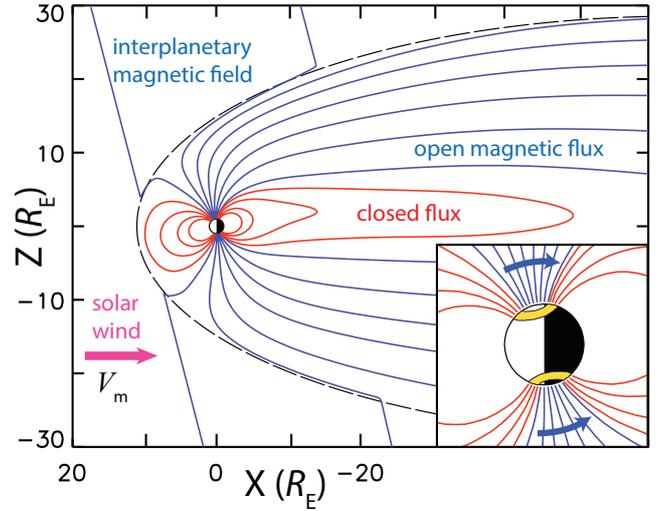}
\caption{
Schematic of the open magnetosphere produced by the Dungey cycle at the Earth.  Closed field lines are shown in red, while open field lines are shown in blue. Adapted from Milan (2009).
}
\label{fig:ms}
\end{figure}

In the case of hot Jupiters, the auroral radio emission is assumed to be generated by a star-planet interaction, mediated either by Alfv\'en waves such as for the sub-Alfv\'enic Io-Jupiter interaction, or via magnetic reconnection as at the Earth.  In the former case, magnetic field lines convecting past the satellite are locally slowed owing to the generation of electric currents in the conductive mantle, forming a steady-state Alfv\'en wave, or Alfv\'en wing, structure propagating away from the satellite.  \cite{Saur:2013dc} considered the Poynting flux radiated away from such Alfv\'en wing structures, based on observations of the Galilean satellites, and computed radiated powers of $\sim$$10^{19}$~W in some cases.  In the latter case, \cite{jardine08a} considered the energy dissipated following reconnection of the planetary and interplanetary field lines, and showed that radio power emitted by such a process would saturate as the orbital distance decreases, owing to the competing effects of increasing stellar wind number density and decreasing magnetospheric size.  However, to date no study has computed the ionospheric plasma flows and currents, and thus the radio power, determined from the resulting ionospheric convection, a process which drives the majority of Earth's auroras.  As shown in Fig.~\ref{fig:ms}, open magnetic flux is created at the dayside magnetopause by reconnection between the planetary and interplanetary magnetic fields, and is then dragged anti-sunward over the poles by the flow of the solar wind to form a long (several thousand Earth radii) magnetotail.  Further reconnection in the tail closes open flux in episodic, energetic events, following which newly-closed flux convects back to the dayside at lower latitudes, completing the process known as the Dungey cycle \citep{dungey61}.  This convection cycle drives a twin-cell vortical flow pattern in the ionosphere, along with an associated magnetospheric current system.  The  component of the current system that flows upward along the magnetic field (associated with downward-precipitating electrons) is responsible for the generation of auroral emissions and the CMI.  In estimating the radio power generated by these processes at exoplanets, the typical procedure is to employ an empirical scaling relation based on observations of bodies in the solar system, known as the `Radiometric Bode's Law', (RBL) which relates incident Poynting or kinetic energy flux to output radio power \citep[e.g.][]{farrell99a,zarka07b}. Extrapolation of the RBL to the estimated input energy fluxes at hot Jupiters (orbiting at typically \ensuremath{\sim}10 stellar radii from of their parent stars) has led to the expectation that next generation radio telescopes may be able to detect such objects \citep{farrell99a, farrell04a, zarka07b, griessmeier07a}.  However, its empirical nature limits how much can be inferred from the RBL.  For example, the radio powers for each planet are assumed to be associated with the solar wind, although for the outer planets the dominant source of power for the auroral current system is the planets' rotation. The radio powers are assumed to scale linearly with the power incident on the dayside of the magnetosphere, although experience at solar wind-driven magnetospheres in the solar system (Earth being the most studied, of course) indicates that the Dungey cycle convection-induced cross-polar cap potential saturates for high values of the motional electric field of the solar wind \citep[e.g.][]{hill:1976a,Siscoe:2002ie, Hairston:2005ei, Kivelson:2008jj}, limiting the power dissipated in the coupled magnetosphere-ionosphere system.  In this paper we thus present calculations of the radio power generated by an Earth-type Dungey cycle at hot Jupiters.  We compute the densities of the magnetosphere-ionosphere coupling currents and the associated precipitating electron energy flux, taking into account the stellar wind conditions and ionospheric conductance at different orbital distances, and polar cap potential saturation.  We show that the radio powers do not increase as quickly with decreasing distance as for the RBL, which leads to lower power values in the region associated with hot Jupiters, but higher powers further out.  We show that young systems are likely to generate higher powers than those of a solar age, such that although detection with LOFAR may prove challenging for systems beyond $\sim$15~pc, many configurations of planets should be detectable with SKA. \\

\nocite{Milan:2009iw}
\section{Theoretical background}
\label{sec:theory}

\subsection{Convection model and field-aligned current}

In this section we present the theoretical background to the problem, and outline the model we employ to estimate the radio powers.  In planetary magnetospheres, CMI-induced radio emissions are beamed from the magnetic field lines at high latitudes, above the auroral zone.  The auroras and CMI are both excited by upward magnetic field-aligned currents (i.e.\ electric currents flowing along a planet's magnetic field lines, away from the planet), which are in general driven by field-aligned voltages that accelerate magnetospheric electrons down the field lines.  Those electrons that are not mirrored then precipitate to the atmosphere and their kinetic energy is dissipated as heat and auroral emissions.  The low-$\beta$, unstable plasma population between the ionosphere and the field-aligned voltage is then favourable for the generation of the CMI, which converts precipitating electron kinetic energy flux to radio power at the rate of $\sim$1\% \citep{wu79a,2006A&ARv..13..229T,zarka98a,lamy:2010aa}.  \\

\begin{figure}
 \noindent\includegraphics[width=20pc]{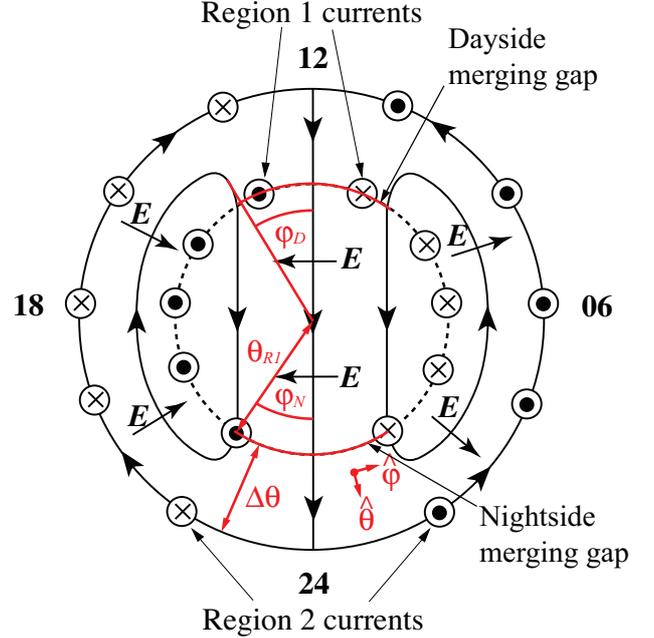} 
\caption{
Schematic of the Dungey cycle flow mapped into the ionosphere, where the arrowed solid lines are the plasma streamlines, the short arrows give the direction of the electric field, and the dashed line in the open-closed field line boundary.  The direction of the field-aligned currents is indicated by the circular symbols, such that the circles with a dot represents upward current, while the circles with a cross indicate downward current. Adapted from Cowley~(2000). 
} 
\label{fig:flows} 
\end{figure}

In an ideal collisionless magnetised plasma, electric currents cannot flow perpendicular to the magnetic field, as described by Alfv\'en's frozen-in theorem.  In a collisional ionosphere, however, such perpendicular currents can occur, the `Pedersen' current flowing parallel to any imposed electric field and the `Hall' current flowing perpendicular to the electric field (i.e. along plasma flow streamlines).  Magnetic field-aligned currents are then a result of current continuity, occuring if there exits a divergence in these field-perpendicular currents.  Such a divergence is the result of the nature of the ionospheric electric field, which is generated by the driving of plasma flows in the ionosphere by some external process.  In the case of a magnetosphere driven by a Dungey-type interaction, the flow pattern is shown schematically in Fig.~\ref{fig:flows}.  The anti-sunward flow of  the solar wind drags newly-opened flux across the open-closed field line boundary (OCFB) at the dayside through a narrow region termed the dayside merging gap. Open flux then flows across the polar cap (PC) as it sinks through the tail lobe toward the equatorial plane.  Reconnection in the tail forms a second, night side, merging gap, whereupon newly closed flux then forms a return flow (RF) back to the dayside at lower latitudes forming the twin-cell convection pattern in Fig.~\ref{fig:flows}.  In the presence of the planet's magnetic field, these ionospheric plasma flows generate, through $\mathbf{E}=-\mathbf{V}\times\mathbf{B}$, roughly-horizontal electric fields in the Pedersen layer of the ionosphere perpendicular to the flow streamlines and the magnetic field (which is near-radial in polar regions).  As discussed above, the divergence of these near-horizontal electric fields requires the presence of field-aligned currents, which form concentric rings known as the Region 1 (R1) current, which flows at the OCFB at co-latitude $\theta_{R1}$, and the Region 2 (R2) current, which flows at the equatorward edge of the return flow region at co-latitude $\theta_{R2}$ as shown in Fig.~\ref{fig:flows}.  These field-aligned currents act to communicate the torque between the magnetosphere and the ionosphere, and the upward R1 and R2 currents together form the auroral oval and the region of CMI-generation.  For a given planet of radius $R_p$, the magnitude of the  field-aligned current density (in A~m$^{-2}$) is dependent on the velocity of the ionospheric plasma flow, characterised by the cross-polar cap convection potential $\Phi_\mathrm{conv}$ induced by reconnection at the dayside magnetopause and in the magnetotail, and the ionospheric Pedersen conductance $\Sigma_P$, such that

\begin{equation}
	j_{\|i}\propto\frac{\Sigma_P\Phi_\mathrm{conv}}{R_p^2}\;\;.
	\label{ipariprop}
\end{equation}

\noindent The details of this relation and its implementation in the present model are deferred to Appendix~\ref{app:deets}, and we now discuss the computation of the convection potential $\Phi_\mathrm{conv}$.\\

\nocite{Cowley:2000km}

\subsection{Cross polar cap potential}
\label{sec:sat}

Dungey cycle convection is driven by reconnection at the nose of the magnetosphere and the tail, and is characterised by a rate of flux transport or, equivalently, a potential induced across the polar cap as discussed above.  The available magnetospheric convection potential $\Phi_m$, is given by the product of the motional electric field of the stellar wind in the rest frame of the planet $E_{sw}$ and the width of the solar wind channel that reconnects, which is in practice some fraction $\chi$ of the magnetopause standoff distance $R_{mp}$, where observationally $\chi\simeq0.5$ \citep{milan04}, which thus also employ here. Hence, we have

\begin{equation}
	\Phi_m=\chi R_\mathrm{mp} E_{sw}\;\;,
	\label{eq:phim}
\end{equation}

\noindent where the magnetopause standoff distance $R_\mathrm{mp}$ is given by

\begin{equation}
	\left(\frac{R_\mathrm{mp}}{R_p}\right)=\left(\frac{k_m^2B_p^2}{2\mu_0(k_{sw}p_{dyn\:sw}+\frac{B_{sw}^2}{2\mu_0}+p_{sw\:th})}\right)^\frac{1}{6}\;\;
	\label{eq:rmp}
\end{equation}

\noindent where $B_p$ is the planetary surface equatorial magnetic field strength, $k_m=2.44$ represents the factor by which the magnetospheric field at the magnetopause is enhanced by magnetopause currents \citep[e.g.][]{mead64a,alexeev05a}, $p_{dyn\:sw}$ is the solar wind dynamic pressure, $k_{sw}=0.88$ for a monatomic stellar wind flow \citep{spreiter70a},  $B_{sw}$ is the interplanetary magnetic field (IMF) strength, and $p_{sw\:th}$ is the (typically negligible) solar wind thermal pressure.  While we have employed standard values for these constants, it is worth noting that the exponent of $1/6$ renders the results insensitive to the exact values. The stellar wind electric field is dependent on the stellar parameters as discussed below in Section~\ref{sec:application}.  \\

This magnetospheric convection potential given by Eq.~\ref{eq:phim} is impressed onto the ionosphere via (to a first approximation) equipotential field lines to become the convection potential $\Phi_\mathrm{conv}$.  The simplest procedure, therefore, would be to take a simple linear dependence of $\Phi_\mathrm{conv}=\Phi_m$, which would then imply a magnetosphere whose convection potential increases linearly with the stellar wind electric field.  This is similar to the assumptions inherent in the RBL.  However, observations of the Earth's magnetosphere (and MHD modelling results) indicate that for high values of $E_{sw}$, the above values of $\Phi_m$ systematically overestimate the actual convection potentials in the polar cap \citep[see e.g.][and references therein]{Hairston:2005ei}.  This phenomenon, which we now briefly review, is known as polar cap potential saturation, discussed initially by \cite{hill:1976a} and developed in a form that expresses the saturation in terms of solar wind parameters by \cite{Siscoe:2002ie} and \cite{Kivelson:2008jj}, the latter two studies approaching the problem from somewhat different physical perspectives.  Initial studies argued that saturation results when the magnetic field associated with the R1 currents (whose sense on the dayside is opposite to the planet's) becomes large enough to reduce the magnetic field at the dayside magnetopause by some significant fraction, thus inhibiting reconnection \citep{Siscoe:2002ie}.  On the other hand, \cite{Kivelson:2008jj} argued that the saturation results since Alfv\'enic perturbations on the open field lines carry signals of the presence of a conducting obstacle (in this case the Pedersen conducting layer of the ionosphere), which are partially reflected from the ionosphere when the solar wind Alfv\'en conductance $\Sigma_A$ is less than the ionospheric Pedersen conductance $\Sigma_P$.  While both physical processes envisaged are distinctly different, the resulting saturation of the convection potential with respect to the solar wind motional electric field is very similar, although for brevity we show results using the model of \cite{Kivelson:2008jj} (hereafter KR), which we now discuss.\\  

The KR model appeals to the fact that the field-aligned component of the magnetosphere-ionosphere coupling currents, which transmit stress between the open field lines and the ionosphere as part of the Dungey cycle, is carried by shear mode Alfv\'en waves.  The Alfv\'en conductance is 

\begin{equation}
	\Sigma_A = \frac{1}{\mu_0 v_A}\;\;,
	\label{eq:siga}
\end{equation}

\noindent where $v_A$ is the Alfv\'en speed given by

\begin{equation}
	v_A = \frac{B}{(\mu_0\rho)^{1/2}}\;\;.
	\label{eq:va}
\end{equation}

\noindent Where the current flows into the ionosphere, the signals are partially reflected owing to the change in impedance between the open field lines and the ionosphere, analogous to the situation for a transmission line for which the impedance of the line does not match that of the load.  The potential transmitted to the ionosphere is

\begin{equation}
	\Phi_\mathrm{conv}=\frac{2\gamma\Phi_m \Sigma_A}{\Sigma_P+\Sigma_A}\;\;,
	\label{eq:phiconvkr}
\end{equation}

\noindent  where the factor $\gamma = (0.1\pi/\chi)$ accounts for the specification of $0.1\pi R_{mp}$ for the width of the interaction channel by \cite{Kivelson:2008jj}.  Saturation occurs when $\Sigma_P >> \Sigma_A$, such that the convection potential tends toward

\begin{equation}
	\Phi_\mathrm{S}=\frac{2\gamma\Phi_m \Sigma_A}{\Sigma_P}\;\;.
	\label{eq:phisatkr}
\end{equation}

\begin{figure}
 \noindent\includegraphics[width=19pc]{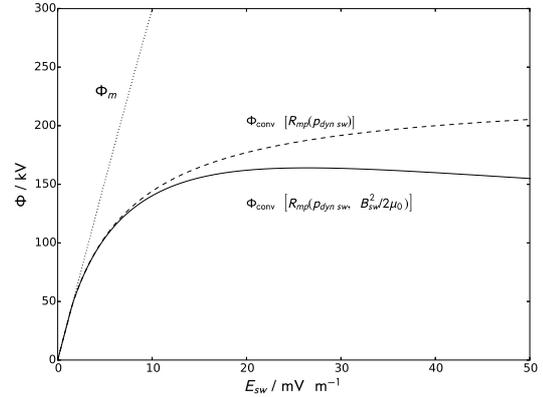}
\caption{
Plot illustrating the saturation of the terrestrial polar cap potential $\phi_\mathrm{conv}$ in kV with solar wind motional electric field $E_{sw}$ in mV\:m$^{-1}$.  The dashed line shows the convection potential values if the IMF magnetic pressure is neglected, the solid line shows the values if this pressure term is included, and the dotted line shows the available magnetospheric convection potential $\phi_m$.  
}
\label{fig:sat} 
\end{figure}

\noindent The saturation effect is illustrated in Fig.~\ref{fig:sat}, in which we plot $\Phi_\mathrm{conv}$ and $\Phi_m$ versus $E_{sw}$ using terrestrial parameters $B_p=31,000\;\mathrm{nT}$, $v_{sw}=400\;\mathrm{km\;s^{-1}}$, $\rho_{sw}=20\times10^{-20}\;\mathrm{kg\;m^{-3}}$ and $\Sigma_P=6\;\mathrm{mho}$. A profile in which the IMF magnetic pressure is not included in the magnetopause pressure balance (equivalent to the case in Fig.~2 of \cite{Kivelson:2008jj}) is shown by the dashed line and the case including this pressure term is shown by the solid line.  It is evident that, while $\Phi_m$ shown by the black dotted line increases linearly with $E_{sw}$, high values of the solar wind motional electric field (which, assuming constant solar wind velocity, is equivalent to low $\Sigma_A$) results in saturation of $\Phi_\mathrm{conv}$ at value of $\Phi_s\simeq230\;\mathrm{kV}$ if IMF magnetic pressure is not included.  With the inclusion of this term, the profile turns over as the magnetopause stand-off distance decreases with increasing IMF strength, and the available convection potential $\Phi_m$ no longer increases linearly. As shown below, the IMF pressure values are not negligible in the hot Jupiter regime, such that we include this term in calculating $R_\mathrm{mp}$.  \\

\subsection{Field-aligned acceleration and energy flux}
\label{sec:fav}

The field-aligned currents computed as above will in most cases be larger than that which can be carried by unaccelerated magnetospheric electrons alone, and must then be driven by a field-aligned voltage.  Specifically, the maximum field-aligned current density that can be carried by an unaccelerated isotropic Maxwellian population is

\begin{equation}
	j_{\|i0} = en\left(\frac{W_{th}}{2\pi m_e}\right)^{1/2}\;\;,
	\label{eq:jpari0}
\end{equation}

\noindent and the corresponding unaccelerated kinetic energy flux is

\begin{equation}
	E_{f0} = 2enW_{th}\left(\frac{W_{th}}{2\pi m_e}\right)^{1/2}\;\;,
	\label{eq:ef0}
\end{equation}

\noindent where $e$, $m_e$, $n$ and $W_{th}$ are the charge, mass, number density and thermal energy of the electron source population, respectively, the latter being equal to equal to $k_BT$, where $k_B$ is Boltzmann's constant and $T$ is the temperature.  We discuss values of these parameters below, but at planets in the solar system, the high latitude magnetospheric electron source population parameters are such that this limiting current is generally much smaller than the field-aligned currents $j_{\|i}$ that are required by the ionospheric flows, such that field-aligned voltages must develop to drive the current. In order to compute the field-aligned voltage, in common with previous works on powerful exoplanetary and ultra-cool dwarf radio emissions \citep{nichols11a,nichols:2012aa,nichols:2012b} we employ Cowley's~(2006) \nocite{cowley06b} relativistic current-voltage relation given by 

\begin{equation}
	\left(\frac{\ensuremath{j_{\|i}}}{\ensuremath{j_{\|i\circ}}}\right)=
	1 + \left(\frac{e\Phi_\|}{W_{th}}\right)+
	\frac{\left(\frac{e\Phi_\|}{W_{th}}\right)^2}{2\left[\left(\frac{m_ec^2}{W_{th}}\right)+1\right]}\;\;,
	\label{eq:phi}
\end{equation}

\noindent where $c$ is the speed of light and $\Phi_\|$ is the minimum voltage required to drive the current $j_{\|i}$ at the ionosphere.  This formulation assumes that the field-aligned voltage is compact and located high enough up the field line, such that the field strength is much less than that in the ionosphere.  For a dipole field, the magnitude of which drops off with the cube of the distance, this assumption is valid beyond a few planetary radii.  The resulting precipitating electron energy flux is 

\begin{equation}
		\left(\frac{E_f}{E_{f0}}\right) = 
		1 + \left(\frac{e\Phi_\|}{W_{th}}\right) + 
			\frac{1}{2}\left(\frac{e\Phi_\|}{W_{th}}\right)^2 +	
			\frac{\left(\frac{e\Phi_\|}{W_{th}}\right)^3}
			{2\left[2\left(\frac{m_ec^2}{W_{th}}\right)+3\right]}\;\;,
	\label{eq:ef}
\end{equation}

\noindent from which the precipitating power for each current, $P_{e\,R1}$ and $P_{e\,R2}$ is obtained by integration over the region of upward current.  In the model, the currents are opposite in the dawn and dusk hemispheres, such that\ 

\begin{equation}
	P_{e\,R1} = \pi R_P^2\,\Delta\theta_{j_{\|i}}\sin\theta_{R1}\int_0^\pi E_{f\,R1}\,d\varphi\;\;,
	\label{eq:per1}
\end{equation}

\noindent and
 
\begin{equation}
	P_{e\,R2} = \pi R_P^2\,\Delta\theta_{j_{\|i}}\sin\theta_{R2}\int_\pi^{2\pi} E_{f\,R2}\,d\varphi\;\;,
	\label{eq:per2}
\end{equation}

\noindent and the total precipitating power is then given by

\begin{equation}
	P_e=P_{e\,R1} + P_{e\,R2}.
	\label{eq:pe}
\end{equation}

\noindent Assuming that we can observe the beam from only one hemisphere at once, and that the electron cyclotron maser instability has a generation efficiency of \ensuremath{\sim}1\%, as discussed above, the total radio power is then given by 

\begin{equation}
	P_r=\frac{P_e}{100}\;\;.  
	\label{eq:pr}
\end{equation}

\noindent and the spectral flux density is finally obtained using

\begin{equation}
	F_r=\frac{P_r}{1.6 s^2 \Delta \nu}\;\;,
	\label{eq:lr}
\end{equation}

\noindent where $\Delta \nu$ is the emission bandwidth, $s$ is the distance to the system from Earth and the emission is assumed to be beamed into 1.6~sr in conformity with Jupiter's DAM and HOM emissions \cite{zarka04a}. The radio emission is generated at the local electron cyclotron frequency, such that the bandwidth is determined by the difference between the field strengths at the ionosphere and the field-aligned voltage, i.e.\ large as discussed above. We thus assume that the bandwidth $\Delta \nu$ is given by the electron cyclotron frequency in the polar ionosphere, i.e.\ 

\begin{equation}
	\Delta\nu=\frac{eB_i}{2\pi m_e}\;\;,
	\label{eq:delnu}
\end{equation}

\noindent an approximation validated by observations of solar system planets \cite{zarka98a}.

\subsection{Application to exoplanets}
\label{sec:application}

\subsubsection{Sun-like star}
\label{sec:sunlike}

\begin{figure*} 
\noindent\includegraphics[width=35pc]{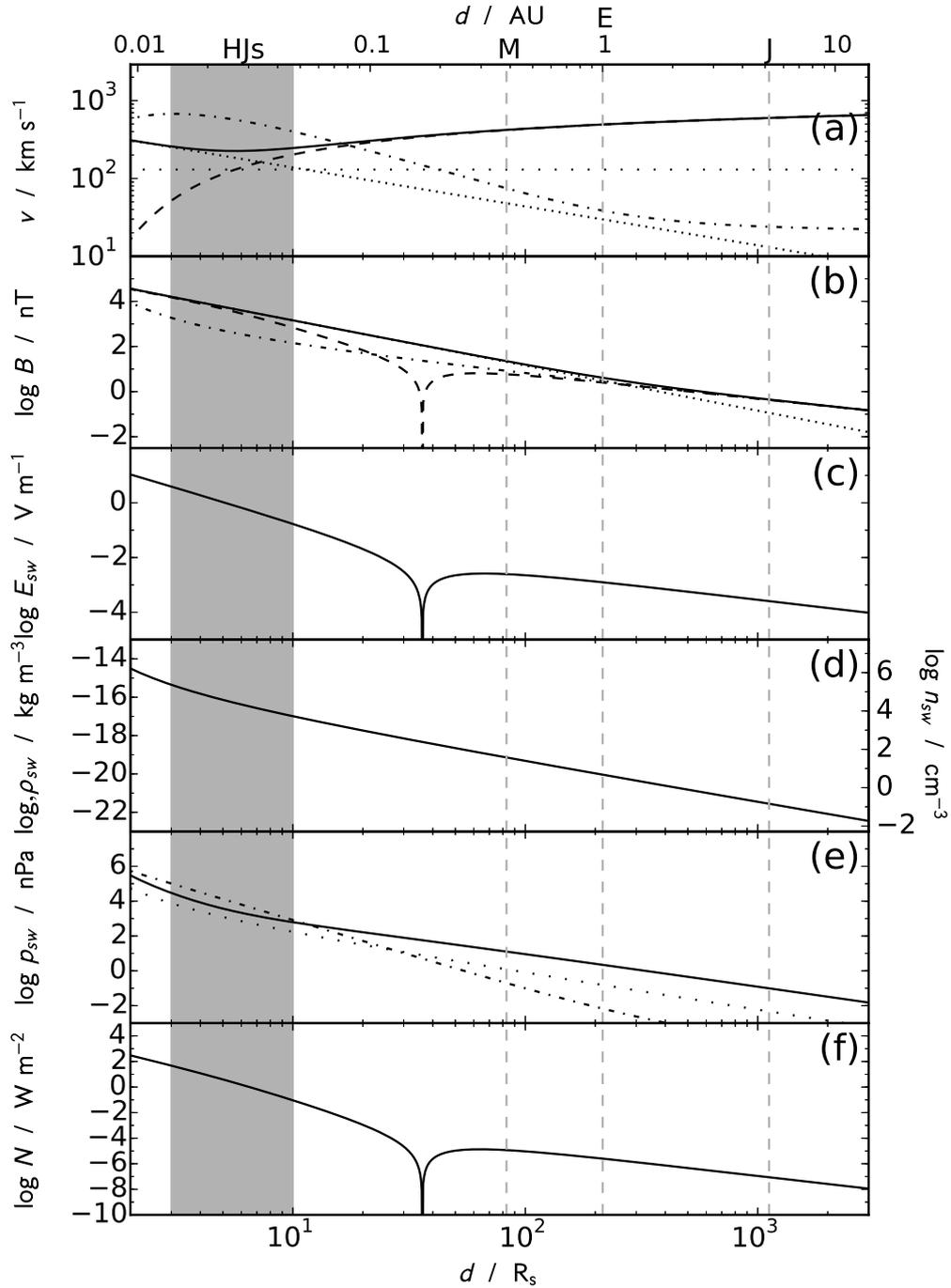} 
\caption{ Plot of the relevant stellar wind parameters versus orbital distance $d$ in stellar radii $\mathrm{R_s}$. Panel (a) shows the stellar wind velocity $V_{sw}$ (dashed line), the Keplerian velocity of a planet in a circular orbit $v_{orb}$ (close-dotted line), and the resultant impinging stellar wind velocity $v_m$ (solid line), along with the stellar wind sound speed $c_s$ (loose-dotted line), and Alfv\'en velocity $v_A$ (dot-dashed line), all in km\:s$^{-1}$. Panel (b) shows the interplanetary magnetic field components, i.e.\ the azimuthal component $B_\varphi$ (dotted line), the radial component $B_r$ (dot-dashed line), the resultant IMF magnitude $|B|$ (solid line), and the component perpendicular to the incident stellar wind velocity $v_\perp$ (dashed line), all in nT. Panel (c) shows the stellar wind electric field $E_{sw}$ in V\.m$^{-1}$. Panel (d) shows the stellar wind mass density $\rho_{sw}$ in kg\:m$^{-3}$ (left axis), and the equivalent number density $n_{sw}$ in cm$^{-3}$ assuming solar average particle mass. Panel (e) shows the solar wind dynamic pressure (dashed line), IMF magnetic field pressure (dot-dashed line) and thermal pressure (loose dotted line) in nPa. Finally, panel (f) shows the stellar wind Poynting flux $N$ in W\:m$^{-2}$. Also shown by the grey bar is the region associated with hot Jupiters, i.e.\ 3-10 stellar radii. The vertical dashed grey lines indicate the orbits of Mercury, Earth, and Jupiter. The top axis indicates the conversion of distance to AU, valid in the case that the stellar radius is equal to the solar radius.} \label{fig:sw} \end{figure*} 

The above formulation in principle applies to any planet with a Dungey cycle-type stellar wind-magnetosphere interaction, and we thus consider here the appropriate parameters for exoplanets orbiting at arbitrary distances, with an emphasis on close-orbiting hot Jupiters.  As discussed above, whereas for the RBL the radio powers are computed as functions of incident kinetic or Poynting flux, in our model the powers are principally functions of the motional electric field of the solar wind, the dynamic pressure of the stellar wind and the Pedersen conductance of the ionosphere, all of which are dependent on further stellar and planetary parameters as described below.  We examine results for both a solar-like stellar wind, and that representative of a young Sun-like star with high mass loss rate and magnetic field strength relative to the Sun.  Considering first the Sun-like stellar wind, the relevant parameters are shown in Fig.~\ref{fig:sw} versus radial distance $d$ normalised by the solar radius $R_s$ (we truncate the inner radial distance of the plot at 2~$\mathrm{R_s}$, being the canonical location of the heliospheric magnetic field `source surface' \citep{Owens:2013hj}). Absolute distances in AU are shown on the top axis for information, although we recognise that in reality the conversion from $\mathrm{R_s}$ to AU depends on the individual star.  Specifically, Fig.~\ref{fig:sw}a shows with the solid line the incident velocity of the solar wind on the magnetosphere $v_m$, which is a function both of the stellar wind speed and the planet's orbital speed.  For simplicity we employ Parker's isothermal solution for the stellar wind speed $v_{sw}$ \citep{Parker:1958tv}, which is fully parameterised by the sound speed $c_s$, and which, as shown by \cite{Cranmer:2004ks} has the closed-form solution

\begin{equation}
	{v_{sw}}^2=	
	\begin{cases}
		  -v_c^2W_0[-D(d)] & \text{if } d \leq d_c \;\;,\\
		  -v_c^2W_{-1}[-D(d)] & \text{if } d \geq d_c \;\;,\\
	  \end{cases}
	\label{eq:vsw}
\end{equation}

\noindent where $W_0$ and $W_{-1}$ are branches of the Lambert $W$ function, $d_c$ is the critical distance at which $v_{sw}$ passes through the sound speed $c_s$, given by

\begin{equation}
	d_c=\frac{GM_s}{2c_s^2}\;\;,
	\label{eq:dc}
\end{equation}

\noindent where $M_s= 1.9891\times 10^{30}$~kg is the solar mass and $D(d)$ is given by

\begin{equation}
	D(d)=\left(\frac{d}{d_c}\right)^{-4}\exp\left[4 \left(1-\frac{d_c}{d}\right)-1\right]\;\;.
	\label{eq:}
\end{equation}

\noindent For the Sun-like wind we employ a sound speed $c_s=130$~km~s$^{-1}$ (which, for a Sun-like average particle mass of $1.92\times10^{-27}$~kg corresponds to a temperature of $\sim$1.18~MK, though note for the present Sun calculation we actually make no assumptions in this regard), yielding a velocity at 1~AU of $\sim$480~km~s$^{-1}$, consistent with observations, and $\sim$50--200~km~s$^{-1}$ in the hot Jupiter region of 3-10~$\mathrm{R_s}$ (indicated by the grey region).  The dotted line indicates the Keplerian speed of a planet in a circular orbit $v_{orb}$, and the solid line is the sum in quadrature of the two, giving the resultant incident stellar wind speed $v_m$.  Note that the two speeds are comparable in the inner region associated with hot Jupiters, and although this will modify the orientation of the magnetosphere with respect to the radial vector, it will not significantly alter the magnetospheric dynamics.  We further show with the loosely dotted and dot-dashed lines the (constant) sound speed $c_s$ and the Alfv\'en speed $v_A$ given by Eq.~\ref{eq:va}.  It is evident that the interaction is everywhere supersonic (modestly so in the hot Jupiter region, with a Mach number of $\sim$2) but becomes sub-Alfv\'enic inside of $\sim$15~$\mathrm{R_S}$, such that Alfv\'en wings will form along the IMF field lines, as discussed by \cite{Saur:2013dc}, effectively shielding the stellar wind motional electric field and is related to KR saturation of the convection potential.  \\

In Fig.~\ref{fig:sw}b we show the IMF components for the Parker Spiral, i.e.\ 

\begin{equation}
	B_r=B_0\left(\frac{d_0}{d}\right)^2\;\;,
	\label{eq:br}
\end{equation}

\noindent and

\begin{equation}
	B_\varphi=B_r \frac{\Omega_s d}{v_{sw}}\;\;,
	\label{eq:}
\end{equation}

\noindent where here $d_0=1R_s$, $B_0$ is the stellar surface field strength (note that we employ the stellar surface here to compare with previous works that consider this parameter; the interplanetary magnetic field is typically considered to be radial at the source surface rather than the solar surface though for our purposes this distinction is not important as we only consider planets outside this radius), and $\Omega_s=2.904\times 10^{-6}$~rad~s$^{-1}$ is the solar angular velocity.  As with \cite{griessmeier07b}, we employ the solar value $B_0=B_{0s}=143,000$~nT (equivalent to 1.43~G), yielding the canonical observed solar minimum value of $B_{sw}=4$~nT at 1~AU.  The dotted and dot-dashed lines indicate the radial and azimuthal components of the magnetic field $B_r$ and $B_\varphi$, the solid line shows the total field $|B|$, and the dot-dashed line indicates the component perpendicular to the stellar wind incidence direction $B_\perp$, i.e. that which gives rise to the motional electric field in the rest frame of the planet, given by

\begin{equation}
	B_\perp = B_{sw}\;\sin\left[\mathrm{arctan} \left(\frac{B_\varphi}{B_r}\right)-\mathrm{arctan} \left(\frac{v_{orb}}{v_{sw}}\right)\right]\;\;.
	\label{eq:bperp}
\end{equation}

\noindent As discussed by \cite{zarka07b}, the `notch' in the vicinity of $\sim$35~$\mathrm{R_s}$ is where the IMF becomes parallel to the incident solar wind velocity, such that in this model the electric field reduces to zero at this point, although it is unlikely that in practice the convection would  reduce to zero, owing to either reconnection on the flanks or convection driven by a viscous interaction.  In the inner region, $B_\perp$ thus varies with distance somewhat faster than does $B_{sw}$, i.e.\ approximately as $B_\perp\propto d^{-17/6}$.  The magnitude of the stellar wind motional electric field $E_{sw}$, shown in Fig.~\ref{fig:sw}c, is then given by $E_{sw}=v_mB_\perp$.  Its value in the hot Jupiter region between 3-10~$\mathrm{R_s}$ is $\sim$0.2-4~$\mathrm{Vm^{-1}}$, i.e.\ roughly two orders of magnitude larger than that typically experienced by the Earth.  The stellar wind mass density $\rho_{sw}$ follows from the stellar wind velocity and the stellar mass loss rate $\dot{M_s}$, and is given by


\begin{equation}
	\rho_{sw}=\frac{\dot{M_s}}{4\pi d^2 v_{sw}}\;\;,
	\label{eq:rhosw}
\end{equation}

\noindent which is shown in Fig.~\ref{fig:sw}d, along with the corresponding number density if the average particle mass were solar.  Here we take the solar value of $\dot{M_s}=2\times 10^{-14}M_s\;\mathrm{yr^{-1}}$, such that the densities in the hot Jupiter region are $\sim$$1-45\times 10^{-17}$~kg~m$^{-3}$, which would correspond to number densities $n_{sw}$ of $\sim$$5-230\times 10^9$~m$^{-3}$ with solar average particle mass.  The stellar wind dynamic pressure, given by 

\begin{equation}
	p_{dyn\:sw}=\rho_{sw} v_m^2\;\;,
	\label{eq:psw}
\end{equation}

\noindent is shown by the solid line in Fig.~\ref{fig:sw}e, along with the solar wind thermal pressure (loose-dotted line) and IMF pressure (dot-dashed line).  The thermal pressure is everywhere negligible compared to the dynamic and magnetic field pressures, which take values between 3 and 10 $\mathrm{R_S}$ of $\sim$$0.6-29.5\times 10^3$~nPa and $\sim$$0.8-100\times 10^3$~nPa, respectively. Thus, in the hot Jupiter region the IMF magnetic field pressure dominates the pressure balance.  Finally, in Fig.~\ref{fig:sw}f we show the magnitude of the stellar wind Poynting flux $N$ given by 

\begin{equation}
	N=E_{sw}B_\perp/\mu_0\;\;,
	\label{eq:poyn}
\end{equation}

\noindent which increases rapidly and has values of $\sim$$0.09-47\;\mathrm{W\;m^{-2}}$ in the hot Jupiter region.  It is this rapid increase in the Poynting flux that has led to the previous suggestions that strongly-driven magnetospheres of hot Jupiters may be detectable using e.g.\ LOFAR.  \\

\begin{figure*} 
\noindent\includegraphics[width=35pc]{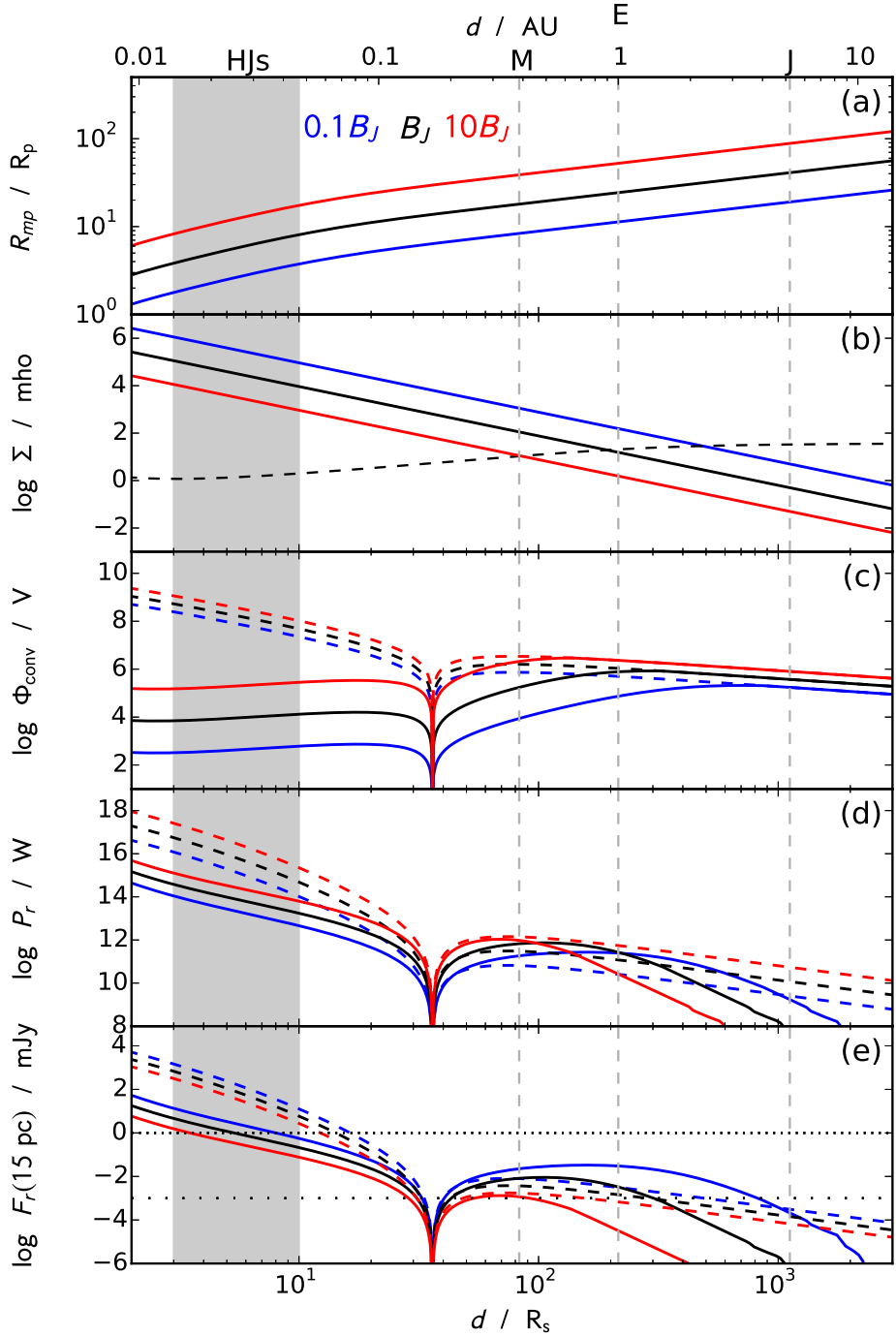}
\caption{
Plot of the planetary parameters versus orbital distance $d$ in stellar radii $\mathrm{R_s}$, for $B_p = 0.1B_J$ (blue lines), $B_J$ (black lines) and $10B_J$ (red lines).  Panel (a) shows the sub-solar magnetosphere standoff distance $R_{mp}$ from Eq.~\ref{eq:rmp} in planetary radii.  Panel (b) shows the ionospheric Pedersen conductance $\Sigma_P$ (solid lines) and the stellar wind Alfv\'en conductance $\Sigma_A$ (dashed line) in mho.  Panel (c) shows the magnetospheric convection potential $\phi_\mathrm{conv}$ (solid lines), along with the available magnetospheric convection potential (dashed lines), in V.  Panel (d) shows the radio power $P_r$ for the saturated case (solid lines), and the RBL (dashed lines) in W.  Panel (e) shows the spectral flux density $F_r$ in mJy assuming a distance of 15~pc, using the same format as in panel (e), except dotted and loose 
-dotted lines show 1~my and 1$\mu$Jy, respectively.  The grey region and vertical dashed lines are as in Fig.~\ref{fig:sw}.}
\label{fig:params} 
\end{figure*}

We now consider the planetary parameters derived from the above stellar wind conditions, as shown in Fig.~\ref{fig:params}, with numerical values of key parameters extracted at 3~$\mathrm{R_S}$ and 10~$\mathrm{R_S}$ given in Table~\ref{tab:results}.  We first show in Fig.~\ref{fig:params}a the size of the magnetosphere computed using Eq.~\ref{eq:rmp}, where we have taken $R_p=R_J$, recognising that there is considerable variation in this parameter.  We consider three values of the planetary magnetic field strength $B_p$, equal to 0.1, 1, and $10B_J$, shown by the blue, red and black lines, respectively.  Under the assumption of constant $v_{sw}$, Eq.~\ref{eq:rmp} yields $R_\mathrm{mp}\propto R_pB_p^{1/3}d^{1/3}$ in the outer region where the solar wind dynamic pressure dominates and $R_\mathrm{mp}\propto R_pB_p^{1/3}d^{2/3}$ in the inner region where the IMF magnetic pressure dominates.  It is thus evident that higher field strengths yield larger magnetospheres, and smaller orbital distances yield smaller magnetosphere size, i.e. $\sim$2-4, $\sim$4-8, and $\sim$8-17~$R_p$ for the three magnetic field strengths, owing to the increased dynamic pressure.  The effect of this is to decrease the width of the channel that is able to reconnect, partially offsetting the increased electric field experienced in this region. \\

Considering now the variation of the Pedersen conductance shown in Fig.~\ref{fig:params}b we employ an expression derived from the jovian value and the modelled Pedersen conductance derived by \cite{koskinen10a} for a hot Jupiter (in particular, HD 209458b), as we now discuss.  First, the conductivity generated by stellar X-ray and EUV (together, XUV) photons introduces both a dependence on the XUV luminosity of the star $L_{XUV}$, such that $\Sigma_P\propto L_{XUV}^{1/2}$, and on radial distance, i.e.\ $\Sigma_P\propto d^{-1}$ \citep[see e.g.][]{nichols11a}.  We take the X-ray luminosity as a proxy for the XUV band as a whole, since X-ray and EUV luminosities are broadly correlated \citep{hodgkin:1994aa}.  Further, the increased scale height of the atmosphere with decreased orbital distance leads to a taller ionosphere, further increasing the conductance over that introduced by increased conductivity alone.  Values of the conductance at Jupiter are not well constrained, although values of order $\sim$0.1--0.5~mho are typically employed (e.g.\ \cite{cowley01,cowley02}), while for HD 209458b orbiting its (assumed Sun-like) star at 0.047~AU, \cite{koskinen10a} computed Pedersen conductances of $9\times 10^3$~mho and $7\times 10^7$~mho for `strong' (i.e.\ $B_p=B_J$) and `weak' (i.e.\ $B_p\simeq0.01B_J$) planetary magnetic field strengths, respectively.  Note that the stronger planetary field yields a lower Pedersen conductance owing to the lower altitude (and thus lower ionisation fraction) of the Pedersen conducting layer, such that canonically $\Sigma_P\propto B_p^{-1}$ \citep{Rassbach:1974ic}. Drawing these various dependences together, we thus employ a power law of the form

\begin{equation}
	\Sigma_{P}= \kappa \left(\frac{d}{1\;\mathrm{AU}}\right)^{\lambda} \left(\frac{B_J}{B_p}\right) \left(\frac{L_{XUV}}{L_{XUV\sun}}\right)^{\mu}\;\;\mathrm{mho}\;\;,
	\label{eq:sigmap}
\end{equation}


\noindent where $\kappa=15.475$, $\lambda=-2.082$, and $\mu=1/2$, such that for a Sun-like star and a Jupiter-like planetary field strength $\Sigma_P=0.5$~mho at $d=5.2$~AU and $\Sigma_P=9\times 10^3$~mho at $d=0.047$~AU, while different stellar and planetary magnetic field values modify the conductance accordingly.  It is worth noting that this expression also yields $\Sigma_P\simeq2.6$~mho for Saturn, consistent with modelled values \citep{moore10a}.  For hot Jupiters, this expression yields $\sim$$9\times 10^4$--$1\times 10^6$, $\sim$$9\times 10^3$--$1\times 10^5$, and $\sim$$9\times 10^2$--$1\times 10^4$~mho for $B_p$ = 0.1, 1, and 10~$\mathrm{B_J}$, respectively. Such conductances are significantly greater than the Alfv\'en conductance shown by the dashed black line in Fig.~\ref{fig:params}b, which for constant $v_{sw}$ would vary as $\Sigma_A\propto d$, although in reality varies in the hot Jupiter region approximately as $\Sigma_A\propto d^{1/2}$ .  The saturation condition of $\Sigma_P>>\Sigma_A$ is thus satisfied in the hot Jupiter region and some way beyond. \\

Turning now to the convection potential shown in Fig.~\ref{fig:params}c, we show the available potential $\Phi_m$ obtained using the simple linear dependence on $E_{sw}$, i.e.\ Eq.~\ref{eq:phim}, with the dashed lines, along with the saturated potentials (solid lines) computed as discussed above.  It is evident that the saturated profiles asymptote to the linear profiles in the outer region, for which the above considerations yield $\phi_m\propto R_pB_p^{1/3}d^{-2/3}$ assuming constant $v_{sw}$ and $B_\varphi$-dominated IMF (i.e.\ $B_\perp \propto d^{-1}$).  In the inner region the saturated profiles diverge to significantly lower values, whilst the linear profiles continue to rise as $E_{sw}$ increases. With $B_{sw}\propto d^{-2}$, $B_\perp \propto d^{-17/6}$ as discussed above, and again assuming constant $v_{sw}$, the available convection potential varies as $\Phi_m \propto R_pB_p^{1/3}d^{-13/6}$, while the saturated potential varies with $\Phi_\mathrm{conv} \propto R_pB_p^{4/3}d^{-(7/3+\lambda)}$ under the same assumptions.  In fact, as $v_{sw}$ is not constant the variation is somewhat less steep, approximately as $\Phi_\mathrm{conv} \propto R_pB_p^{4/3}d^{-(5/3+\lambda)}$. Thus, while the available magnetospheric convection potential increases to values of up to $\sim$20--1100~MV, the saturated potentials decrease to substantially lower values of $\sim$0.3--280~kV, depending on the planetary field strength, in the hot Jupiter region. The convection potential is essentially saturated in the inner region where $\Sigma_P>>\Sigma_A$, though the limiting potential is dependent on $\phi_m$ and thus deviates to lower values in the `notch' region.  In reality the convection potential is unlikely to decrease to zero, owing to contributions from reconnection on the flanks and any viscous interactions.  \\

\begin{table}
\centering
\caption{Table showing numerical values of key parameters for planets with $B_p$ = 0.1, 1, and 10~$\mathrm{B_J}$, each at orbital distances of 3~$R_s$ and 10~$R_s$.  }
\label{tab:results}
\begin{tabular}{l|c|c|c|c|c|c}
\multirow{2}{*}{Property}      			& \multicolumn{3}{c|}{$3R_S$} & \multicolumn{3}{c|}{$10R_S$} \\ \cline{2-7} 
                               			& 0.1 BJ   & BJ    & 10BJ  & 0.1 BJ   & BJ     & 10BJ  \\ \hline
$R_{mp}/R_p$                   			& 1.8      & 3.8   & 8.3   & 3.8      & 8.1    & 17   \\
$\Sigma_P/\mathrm{kmho}$       			& 1129     & 113   & 11    & 92       & 9.2    & 0.9   \\
$\Phi_m/\mathrm{MV}$           			& 249      & 535   & 1153  & 23       & 47     & 105   \\
$\Phi_{\mathrm{conv\,KR}}/\mathrm{kV}$ 	& 0.33     & 7.0   & 151   & 0.61     & 13     & 284   \\
$I_{\mathrm{tot}}/\mathrm{GA}$          & 12       & 26    & 57    & 2.7      & 5.9    & 12.6   \\
$P_{r\,\mathrm{RBL}}/\mathrm{TW}$       & 12,000       & 56,000    & 258,000   & 100     & 480   & 2,200   \\
$P_{r\,\mathrm{KR}}/\mathrm{TW}$        & 110      & 377   & 1262  & 4.6      & 17     & 63    \\
$F_{r\,\mathrm{RBL}}/\mathrm{mJy}$      & 1465      & 680   & 316   & 12       & 5.9    & 2.7   \\
$F_{r\,\mathrm{KR}}/\mathrm{mJy}$       & 13        & 4.6   & 1.5   & 0.6     & 0.2   & 0.08 
\end{tabular}
\end{table}
 
Considering now the radio powers $P_r$, we show in Fig.~\ref{fig:params}d the values computed by the model using the solid lines, along with the powers given by the RBL, shown by the dashed lines for comparison (note that unless otherwise stated, in results that follow we employ the incident Poynting flux for the RBL, rather than the incident kinetic energy flux).  In the absence of knowledge of the plasma population, in computing the powers we take the jovian electron densities and temperatures as fiducial values, and note that the powers would be modified according to Eqs.~\ref{eq:ef0} and \ref{eq:ef} in the event that they differ.  It is apparent that all three profiles exhibit a broadly similar variation, in that (notch region aside) the powers tend to increase for decreased orbital distance.  Specifically, for the RBL results shown by the dotted lines, larger magnetic field strengths yield higher radio powers everywhere owing to greater magnetospheric cross section, such that again assuming constant $v_{sw}$ and $B_\perp \propto d^{-17/6}$ for the inner region yields $P_r\propto R_p^2B_\mathrm{p}^{2/3}d^{-13/3}$, and for the outer region with $B_\perp \propto d^{-1}$ we have $P_r\propto R_p^2B_\mathrm{p}^{2/3}d^{-5/3}$.  This rapid variation in power computed using the RBL in the hot Jupiter region leads to values of $\sim$0.1--260~PW. However, the profiles including convection potential saturation exhibit a somewhat more complex behaviour.  In the outer region where the potential is not saturated, the power varies as $P_r\propto R_p^2B_p^{-4/3}d^{2(\lambda-2/3)}$, assuming constant $v_{sw}$, $B_\varphi$-dominated IMF, and employing the non-relativistic limit of the current-voltage relation (i.e.\ $E_f\propto j_{\|i}^2$).  As the potential saturates, however, the power profiles switch to $P_r\propto R_p^{3/2}B_p^{1/2}d^{-5/2}$, in this case employing the relativistic limit of the current-voltage relation (i.e.\ $E_f\propto j_{\|i}^{3/2}$).  The constant of proportionality for the inner region power law is $\sim$$7\times 10^{15}$.   Note that in this case the power is independent of $\lambda$, i.e.\  the dependence of the Pedersen conductance on radial distance.  Overall then, the radio power values in the hot Jupiter region are $\sim$5--1300~TW, as given in Table~\ref{tab:results}.  While for brevity we do not show details of the plasma flows and currents it is, however, worth noting that for the representative case of a hot Jupiter with magnetic field strength $B_J$ orbiting at 10~$\mathrm{R_S}$ the precipitating electrons are accelerated to $\sim$1~MeV, while energy fluxes peak at $\sim$60~$\mathrm{W\;m^{-2}}$, the total precipitating power is $\sim$1~PW and total power dissipated by Joule heating is $\sim$600~TW. \\

Turning now to the spectral flux densities $F_r$ shown in Fig.~\ref{fig:params}e, it is apparent that, owing to its inverse dependence on the bandwidth and thus the magnetic field strength as in Eq.~\ref{eq:lr} and \ref{eq:delnu}, planets with lower magnetic field strengths exhibit higher flux densities over the whole radial range.  Specifically, power law approximations under the same assumptions as discussed above are $F_r\propto R_p^{2}B_p^{-7/3}d^{2(\lambda-2/3)}$ for the outer region and $F_r\propto R_p^{3/2}B_p^{-1/2}d^{-5/2}$ for the inner.  We have chosen 15~pc as the fiducial distance for which to calculate the spectral flux density, as it is apparent that planets lie on the threshold of LOFAR detectability at this distance, although, as discussed above, the detection threshold of SKA is much lower at 1 $\mu$Jy.  The saturated magnetospheres yield spectral flux densities in the hot Jupiter region of $\sim$0.2-13 mJy, lower field strength and smaller orbital distances yielding higher flux densities, such that at 10~$\mathrm{R_S}$ no planets would be detectable using LOFAR, whereas at 3~$\mathrm{R_S}$ all profiles are above the 1~mJy threshold.  This contrasts significantly with the (undetected) very large flux densities of up to a few thousand Jy given by the RBL in this region. The maximum orbital distances at which these model flux density profiles are greater than the 1~$\mu$Jy detection threshold of SKA are $\sim$3.7, 1.4, and 0.4~AU for $B_p$ = 0.1, 1, and 10~$\mathrm{B_J}$, respectively, comparable with or modestly greater than the values of $\sim$2.2, 1.3, and 0.7~AU for the RBL.  However, the steeper gradients of the unsaturated regions of the flux density profiles compared with the RBL in the outer region are such that the modelled flux densities are, for the weaker planetary fields, considerably larger than the RBL results.  For example, for a planet with $B_p$ = 0.1~$\mathrm{B_J}$ orbiting at 1~AU, the modelled flux density is $\sim$30~$\mu$Jy, compared with the RBL's barely-detectable $\sim$3~$\mu$Jy.  \\
 

\subsubsection{Young Sun-like star}
\label{sec:youngsun}

It has been suggested, using the kinetic RBL, that young, fast-rotating stars possessing hot, fast stellar winds with high mass loss rate are likely to produce brighter emissions owing to greater impinging energy fluxes on the magnetospheres of planets \citep{griessmeier07b}.  Here, we thus consider the powers computed using our model for planets orbiting a young main-sequence ($\sim$1~Gy) Sun-like star.  To estimate the stellar wind properties, we employ relations which provide the expected variation of the key solar parameters with age, as determined by observations of solar analogues \citep[see e.g. the review by][]{Gudel:2007gx}.  We first determine the rotation period $P$ in days using the relation given by \cite{1994ASPC...64..399D}, i.e.\ 

\begin{equation}
	P = 0.21 t_6^{0.57}\;\;,
	\label{eq:pdays}
\end{equation}

\noindent  where $t_6$ is the age of the star in My since arriving on the zero-age main sequence, yielding $P\simeq10.8$ days.  From this we compute the X-ray luminosity $L_X$ in erg s$^{-1}$ using the relation of \cite{Gudel:1997du}, given by

\begin{equation}
	L_X = 10^{31.05} P^{-2.64}\;\;,
	\label{eq:lxp}
\end{equation}

\noindent which yields $L_X=10^{28.32}$~erg s$^{-1}$, i.e.\ a factor of $\sim$9.4 larger than the mean solar value of $L_{Xs}=10^{27.35}$ as given by \cite{judge:2003aa}.  The X-ray luminosity is then used to estimate a number of other parameters as follows.  The coronal (and, under the isothermal assumption, stellar wind) temperature $T_{sw}$ in MK using the relation given by \cite{Gudel:2007gx}, i.e.\ 

\begin{equation}
	T_{sw} = \left(\frac{L_X}{1.61\times 10^{26}}\right)^{0.247}\;\;,
	\label{eq:tsw}
\end{equation}

\noindent which gives $T_{sw}=3.3$~MK and thus, for solar wind composition, a stellar wind sound speed of $\sim$219~km s$^{-1}$.  This temperature is within the range of $\sim$1-10~MK observed in solar analogues. The stellar wind mass loss is calculated from $L_X$ via the relation given by \cite{wood05a}, i.e.\ 

\begin{equation}
	\dot{M}=\dot{M_s}\left(\frac{L_X}{L_{Xs}}\right)^{1.34}\;\;,
	\label{eq:}
\end{equation}

\noindent such that $\dot{M}=4\times 10^{-13}\;M_s\,\mathrm{yr^{-1}}$.  We finally assume that the stellar surface field strength $B_0$ is, for fixed $R_s$ and field geometry, proportional to the total stellar magnetic flux, such that we determine $B_0$ from $L_X$ using the relation of \cite{Pevtsov:2003el}, i.e.\ 

\begin{equation}
	B_0=B_{0s}\left(\frac{L_X}{L_{Xs}}\right)^{0.885}\;\;,
	\label{eq:}
\end{equation}

\noindent which yields $B_0=1.04\times 10^6$~nT (equivalent to 10.4~G).  \\ 

\begin{figure*} 
\noindent\includegraphics[width=35pc]{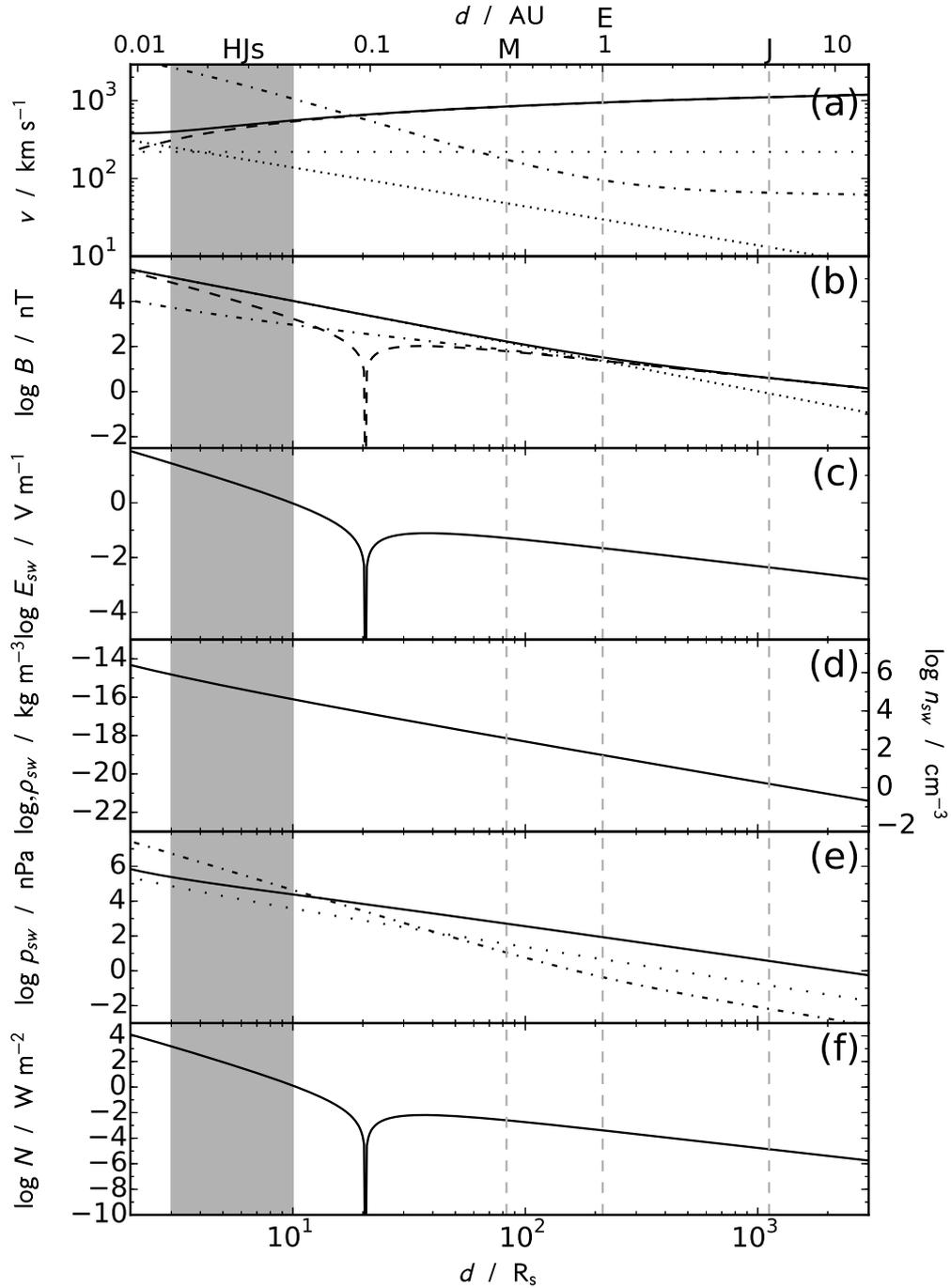}
\caption{
As Fig.~\ref{fig:sw} but for a young Sun-like star. 
}
\label{fig:swy} 
\end{figure*}

We thus show in Fig.~\ref{fig:swy} the stellar wind parameters of a young Sun-like star versus radial distance, in the same format as Fig.~\ref{fig:sw}.  The velocities shown in Fig.~\ref{fig:swy}a are $\sim$310--540~km~s$^{-1}$ in the hot Jupiter region, i.e.\ a factor of $\sim$3--6 higher than for the present Sun.  Hence, while the Alfv\'en speed is also increased, the interaction becomes super-Alfv\'enic outside of $\sim$18~$\mathrm{R_S}$.  The perpendicular magnetic field shown in Fig.~\ref{fig:swy}b is $\sim$72,000--2,200~nT in the hot Jupiter region, i.e. a factor of $\sim$3--4 greater than for the present Sun.  Overall, then, the stellar wind electric field shown in Fig.~\ref{fig:swy}c, which takes values of $\sim$1--29~$\mathrm{Vm^{-1}}$, i.e.\ a factor of $\sim$7 higher than for the current Sun.  The mass densities shown in Fig.~\ref{fig:swy}d are $\sim$$8-152\times 10^{-17}$~kg~m$^{-3}$, which corresponding to number densities of $\sim$$4-79\times 10^{10}$~m$^{-3}$ with solar average particle mass.  The solar wind dynamic pressure shown in Fig.~\ref{fig:swy}e is $\sim$$2.4-24\times 10^4$~nPa, i.e.\ $\sim$8--39 times that of the present Sun in the hot Jupiter region, while the IMF magnetic field pressure is $\sim$$4.3-534\times 10^4$~nPa, i.e.\ a factor of $\sim$52 higher than the present Sun.  Finally, the Poynting flux shown in Fig.~\ref{fig:swy}f is $\sim$2--1626~$\mathrm{W\;m^{-2}}$ in the hot Jupiter region, i.e.\ a factor of $\sim$24--34 greater than for the present Sun. \\

\begin{figure*} 
\noindent\includegraphics[width=35pc]{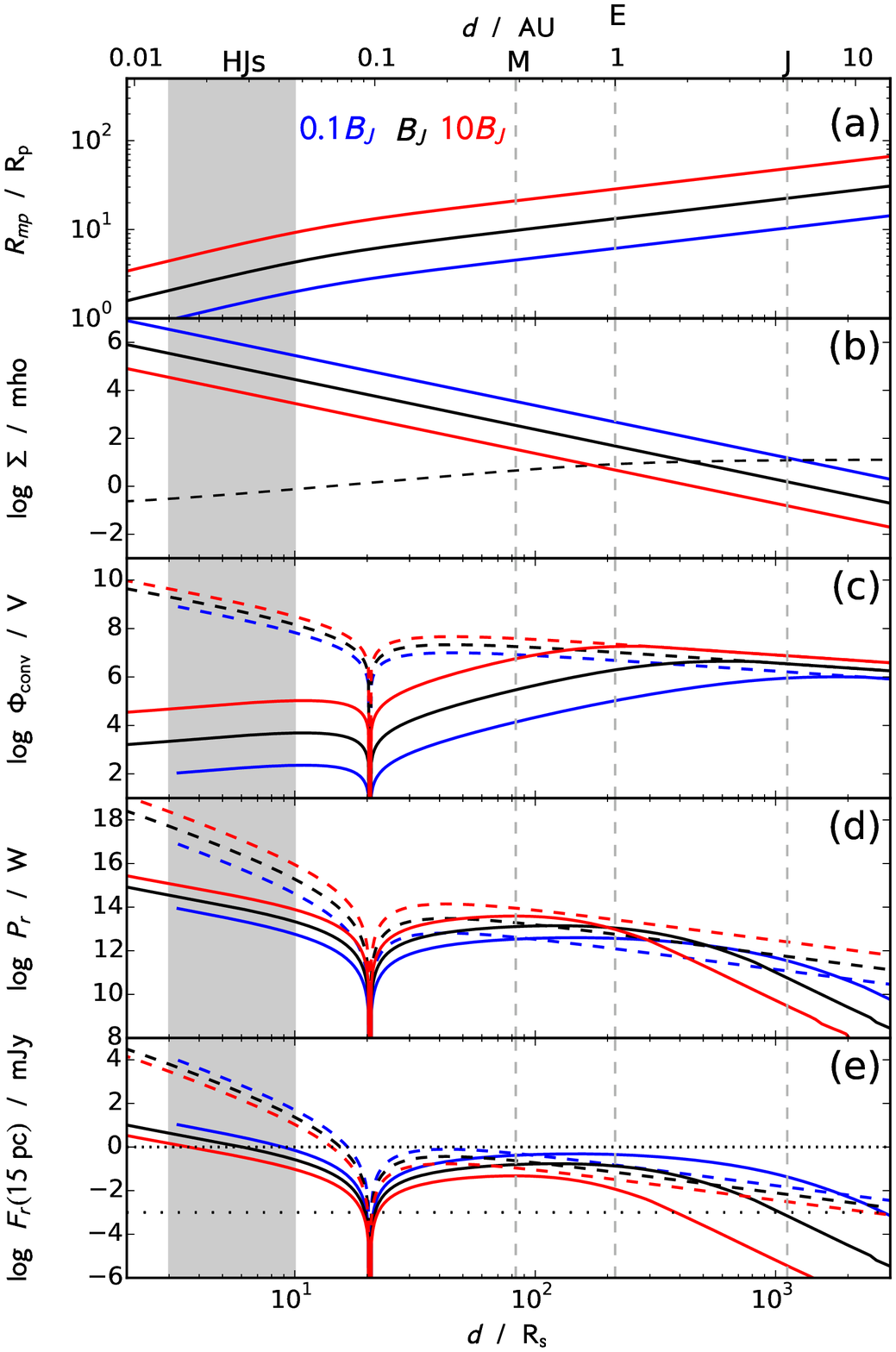}
\caption{
As Fig.~\ref{fig:params} but for a young Sun-like star. 
}
\label{fig:paramsy} 
\end{figure*}

Turning then to the planetary parameters determined from the above stellar wind characteristics, we show profiles in Fig.~\ref{fig:paramsy} in the same format as Fig.~\ref{fig:params} and give numerical values in Table~\ref{tab:resultsy}.  As shown in Fig.~\ref{fig:paramsy}a, the higher dynamic pressure yields smaller magnetospheres for a given magnetic field strength than for the present Sun, i.e.\ $\sim$0.96--2, 2--4, and 4--9~$\mathrm{R_p}$.  Note that for the 0.1 $10B_J$ case, the magnetopause radius becomes less than the planetary radius at $\sim$3.3$\mathrm{R_S}$, such that in the panels below, blue profiles which depend on this parameter are truncated at this distance.  Considering the Pedersen conductance, the values are a factor of $\sim$3 larger those for the present Sun, at $\sim$280-3500~kmho, $\sim$28-350~kmho, and $\sim$2.8-35~kmho for $B_p$ = 0.1, 1, and 10~$\mathrm{B_J}$, respectively.  Although the available magnetospheric convection potentials are somewhat larger than for the present Sun, reaching almost $\sim$5~GV at 3~$\mathrm{R_S}$ for $B_p$ = 10~$\mathrm{B_J}$, the saturated potentials are decreased by a factor of $\sim$0.3, with values of $\sim$0.1--140~kV, depending on the planetary field strength, in the hot Jupiter region.  Thus, the ratio between the two cases of the combined parameter $(\Sigma_P\Phi_S)$ is $\sim$1, and the ionospheric currents and radio powers and flux densities are essentially unchanged in the saturated region from those of the present Sun.  Hence, the flux densities in the hot Jupiter region are generally 2 orders of magnitude below those for the RBL, which reaches few thousand mJy at 3~$R_s$.  The powers are, however greater than the present Sun values in the outer region where the convection potential is not saturated.   The maximum orbital distances at which these flux densities exceed the detection threshold of SKA are $\sim$13, 5, and 2~AU for $B_p$ = 0.1, 1, and 10~$\mathrm{B_J}$, respectively, somewhat less than the distances of $\sim$33, 19, and 11~AU for the RBL owing to the steeper gradient in the outer region. However, the flux densities of the 0.1~$\mathrm{B_J}$ case are greater than the RBL values between $\sim$0.4-8.6~AU and for a planet orbiting at 1~AU, the flux densities are $\sim$3-150~$\mu$Jy, which should be detectable with SKA.

\begin{table}
\centering
\caption{As for Table~\ref{tab:results} but for a young Sun-like star.}
\label{tab:resultsy}
\begin{tabular}{l|c|c|c|c|c|c}
\multirow{2}{*}{Property}                & \multicolumn{3}{c|}{$3R_S$} & \multicolumn{3}{c|}{$10R_S$} \\ \cline{2-7} 
                                         & 0.1 BJ   & BJ     & 10BJ  & 0.1 BJ   & BJ    & 10BJ   \\ \hline
$R_{mp}/R_p$                             & 0.96     & 2.1    & 4.4   & 2.0      & 4.3   & 9.3   \\
$\Sigma_P/\mathrm{kmho}$                 & 3469     & 347    & 35    & 283      & 28    & 2.8   \\
$\Phi_m/\mathrm{MV}$                     & 973      & 2097   & 4518  & 88       & 189   & 406  \\
$\Phi_{\mathrm{conv\,KR}}/\mathrm{kV}$   & 0.11     & 2.3    & 49    & 0.29     & 6.3   & 136    \\
$I_{\mathrm{tot}}/\mathrm{GA}$           & 42       & 91     & 196   & 9.7      & 21    & 45   \\
$P_{r\,\mathrm{RBL}}/\mathrm{PW}$        & 119      & 553    & 2564  & 0.69     & 3.2   & 15   \\
$P_{r\,\mathrm{KR}}/\mathrm{TW}$         & 109      & 376    & 1258  & 9.0      & 33    & 119   \\
$F_{r\,\mathrm{RBL}}/\mathrm{mJy}$       & 14,547   & 6752   & 3134  & 84       & 39    & 18   \\
$F_{r\,\mathrm{KR}}/\mathrm{mJy}$        & 13       & 4.6    & 1.5   & 1.1      & 0.41  & 0.14 
\end{tabular}
\end{table}

\section{Discussion and summary}

The radio powers discussed here are the first to be computed for exoplanets assuming a Dungey-type stellar wind-planet interaction, resulting from magnetospheric convection driven by magnetic reconnection.  They are also the first to be computed considering polar cap potential saturation, which is known to occur at Earth when the magnetosphere is subject to high values of the solar wind motional electric field.  We have determined the powers and flux densities at the representative distance of 15~pc for planets orbiting a Sun-like star and a young 1~Gy Sun-like star.  We have employed the \cite{Kivelson:2008jj} model of polar cap potential saturation, such that saturation occurs when the ionospheric Pedersen conductance is substantially larger than the interplanetary Alfv\'en conductance, and signals propagating into the ionosphere are partially reflected.  The resulting powers are dependent on the available magnetospheric convection potential $\phi_m$ and thus decrease to zero where the IMF becomes aligned with the incident stellar wind velocity.  In reality, the convection potential is unlikely to actually decrease to zero owing to natural variations from the Parker spiral direction, reconnection along the flanks, and viscous interactions.  We have further produced power law approximations to the flux densities that are applicable in the hot Jupiter region with saturated cross polar cap potential, and for the unsaturated profiles in the region further out.   We have shown that the radio powers and flux densities broadly increase with decreasing radial distance, though more slowly in the inner region where the convection potentials are saturated than further out, in contrast to the RBL powers which, `notch' region aside, increase more quickly with decreasing distance.  The saturated profiles also increase with magnetic field strength, in constrast with the unsaturated regime, though the flux densities decrease with field strength everywhere owing to the dependence of the bandwidth on the electron cyclotron frequency at the ionosphere. \\

For a Sun-like star, the computed radio powers for the hot Jupiter region are $\sim$5--1300~TW, roughly two orders of magnitude below those for the RBL. The flux densities are $\sim$0.6~mJy for a field strength of 0.1~$B_J$ at 10~$R_s$, increasing to $\sim$13~mJy at 3~$R_s$. Such fluxes are $\sim$1-2 orders of magnitude below those for the RBL, which thus presents a significant overestimation of the detectability of these exoplanets. At further distances the powers are everywhere less than the detection threshold for LOFAR, but are greater than 1~$\mu$Jy out to $\sim$0.4-3.7~AU depending on the planetary field strength. For a planet with $B_p=0.1~B_J$ orbiting at 1~AU, the flux density is up to $\sim$30~$\mu$Jy, which may be detectable with SKA. For a young Sun-like star, we find that, while the powers estimated by RBL are increased by a factor of $\sim$10, for our model the decreased saturation potential and increased Pedersen conductance provide essentially identical competing effects on the powers, which are thus almost identical to those of the present-day Sun. In the outer unsaturated region, however, the powers the young system are increased over the Sun-like star, and the flux densities are above the SKA detection threshold out to $\sim$2-13~AU, depending on the field strength. Specifically, for planets orbiting at 1~AU, the flux densities are $\sim$3-150$\mu$Jy, depending on the field strength.   \\

As part of the model, we compute the flows and currents in the ionosphere.  A key parameter is the energy of the precipitating electrons and the energy flux. Our results indicate that, for a hot Jupiter with $B_p=B_J$ orbiting at 10~$R_S$, the precipitating auroral electron energies are around $\sim$1~MeV, carrying an energy flux of a few tens of $\mathrm{W\;m^{-2}}$, for a total precipitating power of $\sim$1~PW into the polar atmosphere.  This is a significant energy source whose implications should be considered in atmospheric circulation models for hot Jupiters.  Joule heating from the ionospheric Pedersen currents will then form a further source of heating of the upper atmosphere, as has been suggested e.g.\ by \cite{2013ApJ...765L..25B} and \cite{Cohen:2014eb}.  Our model yields total Joule heating of $\sim$600~TW for the planet at 10~$R_s$, i.e.\ lower than the estimate of \cite{2013ApJ...765L..25B} by several orders of magnitude, though our magnetospheric convection is saturated and cannot dissipate the total available incident solar wind power.  \\

There are some limitations to the model presented here.  The simple steady-state representation of the dynamics of convection in a magnetosphere does not take into account the significant bursty nature of the process, as is evidenced by the sub-storm cycle at Earth \citep{Russell:1973gx}.  Nightside reconnection in particular is bursty, and energy is built up and stored in the magnetotail magnetic field until a burst of reconnection closes a substantial quantity of open flux in a short interval of time, resulting in expanded and brightened auroral emission for a short period of time (an hour or so at Earth). Typical terrestrial substorms occur with frequencies of $\sim$3 hours \citep{Borovsky:1993ks}, releasing $\sim$160\% of the energy in the 2~h post-onset than in the preceding 2~h \cite{Newell:2001fr}. Such bursty behaviour may significantly increase the detectability of auroral radio emissions from exoplanets above those considered by the present model, at the cost of limited temporal opportunity for dxetection.  This inherent bursty nature of the process is in addition to the variability expected via variation of the stellar wind parameters with time and stellar longitude \citep[e.g.][]{2015MNRAS.450.4323S}.  Further, the size of the polar cap at any one time is determined by the quantity of open flux in the tail, which changes significantly over the course of the substorm cycle.  We have taken a polar cap radius of 15$^\circ{}$ in conformity with observations of the typical polar cap size in the solar system, although we note that MHD models of hot Jupiter magnetospheres indicate that the polar cap radii may be significantly larger. Tests indicate that taking a polar cap radius of $\sim$45$^\circ{}$ raise the emitted powers from those presented here by approximately a factor of 2. The present model does not consider any convection potential driven by a viscous interaction at the magnetopause boundary \citep{1961CaJPh..39.1433A}, and the effects of such a process should be examined in future works.  Further, the radio powers would be modified from those presented here if parameters of the high latitude electron source population differ from those assumed here, and indeed any observations of exoplanetary radio emissions will act as a probe for these parameters.  Further, we have assumed a constant ionospheric conductance, which would not be the case for strongly-irradiated hot Jupiters, for which the ionospheric currents would be confined to the dayside, and the feedback on the ionospheric convection should be examined using more complex MHD models.   While we have considered the effects on the radio emissions of the parameters of stars of different ages, we have not examined any corresponding changes in the intrinsic planetary parameters over a several Gyr timespan.  Finally, we have not considered interplay with the flows and currents arising from planetary rotation and internal plasma sources \citep{nichols11a, nichols:2012aa}, which is likely to be a factor for Jupiter-like planets orbiting outside the tidal locking radii, and which should be examined in future using MHD models.

\section*{Acknowledgements}

JDN was supported by an STFC Advanced Fellowship (ST/I004084/1).  SEM was supported by STFC Grant ST/K001000/1.




\bibliographystyle{mnras}
\bibliography{/Users/jdn/Documents/Papers/exprefs}

%
%
\appendix

\section{Details of the convection model}
\label{app:deets}

We employ a simple, extensively used and validated analytic model of ionospheric convection, originally developed to model the plasma flows and currents in and around the expanding and contracting polar cap of Earth \cite[e.g.][]{Siscoe:1985kt,Freeman:1988eo,Freeman:2003ib,Milan:2012fr,milan:2013a}.  The details of the model are given e.g.\ by \cite{milan:2013a}, but briefly, the model assumes that the planet is a sphere of radius $R_p$, such that positions in the ionosphere are given by co-latitude $\theta$ and azimuth $\varphi$, the latter defined such that $\varphi=0$ is oriented toward midnight and $\varphi$ increases in the direction of planetary rotation for the case of a planet with a magnetic moment of the same sense as that of the Earth, i.e.\ southward.  The ionospheric electric field $\mathbf{E}(\varphi,\theta)=E_\varphi \boldsymbol{\hat{\varphi}}+E_\theta \boldsymbol{\hat{\theta}}$ is described by the gradient of a scalar potential $\Phi$, such that $\mathbf{E}=-\nabla\Phi$.  The field-perpendicular current $\mathbf{j_\perp}$ is related to the ionospheric electric field by

\begin{equation}
	\mathbf{j_\perp}=\Sigma_P\mathbf{E}+\Sigma_H\mathbf{\hat{B}}\times\mathbf{E}\;\;,
	\label{eq:jperp}
\end{equation}

\noindent where $\Sigma_P$ and $\Sigma_H$ are the height-integrated Pedersen and Hall conductances, respectively, and $\mathbf{\hat{B}}$ is the unit vector of the magnetic field.  The divergence of the field-perpendicular current yields the field-aligned current intensity, i.e. current per unit azimuthal distance in $\mathrm{A\;m^{-1}}$, at the top of the ionosphere $i_{\|i}$ given by

\begin{equation}
	i_{\|i}=\nabla\cdot\mathbf{j_\perp}=\Sigma_P\nabla^2\Phi+\nabla\Phi\cdot\nabla\Sigma_P+(\nabla\Phi\times\mathbf{\hat{B}})\cdot\nabla\Sigma_H\;\;.
	\label{eq:jpari}
\end{equation}

\noindent In general, the conductances $\Sigma_P$ and $\Sigma_H$ are spatially variable, modified locally by e.g.\ photoionisation and the precipitating electron energy flux, but in the light of the lack of detailed models of the ionospheres of strongly-irradiated hot Jupiters, we simply take the conductances to be equal and uniform across the planet's surface, with values computed as discussed further below.  \cite{milan:2013a} showed that, with the form of the electric potential for the model (given by their Table 1), the R1 and R2 field-aligned current intensities, which flow at co-latitudes $\theta_{R1}$ and $\theta_{R2}=\theta_{R1}+\Delta\theta$, respectively, are then given by

\begin{equation}
	i_{\|i\,R1}=\frac{\Sigma_P}{R_p\sin\theta_{R1}}\sum_{m=1}^Ns_mm\,\sin m\varphi\left[\,\mathrm{coth}\,m(\Theta_{R1}-\Theta_{R2})-1\right]\;\;,
	\label{eq:jr1}
\end{equation}

\noindent and

\begin{equation}
	i_{\|i\,R2}=\frac{\Sigma_P}{R_p\sin\theta_{R2}}\sum_{m=1}^Ns_mm\,\sin m\varphi\,\mathrm{csch}\,m(\Theta_{R1}-\Theta_{R2})\;\;,
	\label{eq:jr2}
\end{equation}

\noindent where $\Theta=\ln\tan\frac{1}{2}\theta$ and $s_m$ is given by

\begin{equation}
	s_m=-\frac{1}{m^2\pi}\left[(-1)^m \frac{\Phi_D\sin m\varphi_D}{\varphi_D}-\frac{\Phi_N\sin m\varphi_N}{\varphi_N}\right]\;\;,
	\label{eq:sm}
\end{equation}

\noindent where $\varphi_{D,N}$ are the angular half-widths of the day- and nightside merging gaps and $\Phi_{D,N}$ are the day- and nightside reconnection voltages, related to the rate of flux transport through the merging gaps via Faraday's Law.  The typical polar cap size at planets in the solar system is $\sim$15-20$^\circ{}$ \citep[e.g.][]{Iijima:1976iv, Jinks:2014dh}, such that  here we take $\theta_{R1}=15$$^\circ{}$, along with $\Delta\theta=10$$^\circ{}$ and $\varphi_{D}=\varphi_{N}=30$$^\circ{}$ following \cite{milan:2013a}.  At any one time the day- and nightside reconnection voltages are in general different, indicating differing rates of dayside and nightside reconnection, but are identical in the steady state and when averaged over many convection cycles, and in which case are parameterised by a single cross-polar cap potential associated with the convection $\Phi_D=\Phi_N=\Phi_\mathrm{conv}$. We discuss the calculation of the values of $\Phi_\mathrm{conv}$ in Section~\ref{sec:sat} below. The sum over $m$ can in principle be taken to any arbitrary $N$, and is taken by \cite{milan:2013a} up to $N=20$, which we thus also employ here.  The currents given by Eqs.~\ref{eq:jr1} and \ref{eq:jr2} are formally assumed to infinitely thin sheets, although computing the precipitating electron energy flux requires the current density $j_{\|i}$ in $\mathrm{A\;m^{-2}}$, and we thus assume that these currents form thin annuli of small but finite thickness $\Delta \theta_{j_{\|i}}$, with uniform latitudinal distributions, such that\ 

\begin{equation}
	j_{\|i\,R1}=\frac{i_{\|i\,R1}}{\Delta \theta_{j_{\|i}} R_p}\;\;,
	\label{eq:jpir1}
\end{equation}

\noindent and

\begin{equation}
	j_{\|i\,R2}=\frac{i_{\|i\,R2}}{\Delta \theta_{j_{\|i}} R_p }\;\;,
	\label{eq:jpir1}
\end{equation}

\noindent and in conformity with observations at Earth, Jupiter and Saturn, we take the thickness $\Delta \theta_{j_{\|i}}=1^\circ{}$.  We note that our results are not strongly dependent on realistic choices of this width.  \\

%
%

\bsp	
\label{lastpage}
\end{document}